\documentstyle[prl,aps]{revtex}
\newcommand{\BE}{\begin{equation}}
\newcommand{\EE}{\end{equation}}
\newcommand{\BA}{\begin{eqnarray}}
\newcommand{\EA}{\end{eqnarray}}
\begin{document}
\draft
\twocolumn[\hsize\textwidth\columnwidth\hsize\csname@twocolumnfalse\endcsname

\title {Localize energy in random media: A new phase state}
\author{Zhen Ye and Kang-Xin Wang}
\address{Wave Phenomena Laboratory, Department of Physics, National Central University, Chungli, Taiwan 32054}

\date{\today} \maketitle

\begin{abstract}

This Letter reports a new wave phenomenon. We show that wave
energy can be stored in random media. Associated with such energy
storage is a global collective behavior. The features depicted are
so general that they may be observed in many random systems.
\end{abstract}
]

A major task in physics research is to search for common
principles behind various nature processes. A number of greatest
advances in the last century include understanding of how the
manifestation of many body interactions may lead to several
important ubiquitous phenomena including superconductivity,
quantum Hall effect, screening effect in plasma, and the Kondo
effects\cite{Mahan}. It becomes a daily experience that
macroscopic objects arise from interaction between individuals of
microscopic scale in a many body system\cite{Umezawa}.

In this Letter, we will show a new general wave phenomenon in
certain many body systems. Namely, wave energy can be stored in a
number of different disordered systems. When this occurs, a new
phase state emerges.

Up to date, the fundamental equation describing an arbitrary many
body system can be more or less written as\BE {\cal L}_i(\partial)
\psi_i = F_i + \sum_{j=1,j\neq i}^N \hat{T}_{ij} [\psi_j],
\label{eq:1}\EE in which ${\cal L}$ is called the
divisor\cite{Umezawa}, $F_i$ denotes the external stimulation,
$\hat{T}_{ij}$ is an operator that can be of temporal, spatial or
both nature, depicting the interaction between consisting bodies.
In the linear approximation, the operator $\hat{T}$ is represented
by a wave propagator. For example, in the Yang-Feldman
representation of field equations, $\hat{T}$ is the usual ferminic
or bosonic two-point Green's function. In Eq.~(\ref{eq:1}), $i$ is
the index of the individuals in the system of total $N$ bodies.

So far, there are two known types of divisor\cite{Umezawa}. Type I
divisor usually refers to fermionic systems and is written as \BE
{\cal L(\partial)} = i\hbar \frac{\partial}{\partial t} -
\epsilon(\nabla),\label{eq:2}\EE Type II divisor refers to
boson-like systems and is written as \BE {\cal L(\partial)} =
\frac{1}{c^2}\frac{\partial^2}{\partial t^2} - \omega^2(\nabla).
\label{eq:3} \EE In these, $\epsilon(\nabla)$ and $\omega(\nabla)$
are the operators yielding the energy spectra.

Equation (\ref{eq:1}) accommodates a variety of problems of great
interest. To name a few, this includes the vibration of the
lattice formed by atoms, spin arrays, and electrons in lattices.
Interested readers may refer to \cite{Ziman} for more examples.
Eq.~(\ref{eq:1}) can also describe multiple scattering of
classical waves such as acoustic propagation in air-filled
bubbles\cite{cjp} and cylinders\cite{Ye} and electromagnetic waves
in electric dipolar systems\cite{dipole}. In these systems, the
interaction between constituents is mediated by waves which
themselves are described by Type II equations.

Here we show some new generic features with Type II systems.
Consider wave propagation in a Type II system. The propagating
wave, which can be either acoustic or electromangetic, will be
interfered by scattered waves from each constituent being
stimulated by the incident wave. The energy flow in the system is
$\vec{J} \sim \mbox{Re}[\psi(-i)\nabla\psi]$. Writing the field as
$\psi = |\psi| e^{i\theta}$, the current becomes
$|\psi|^2\nabla\theta$, a version of Meissner equation. It is
clear that when $\theta$ is constant while $|\psi| \neq 0$, the
flow stops and the energy must be localized in space, hinting at
that a phase transition occurs and modes are condensed in the real
space. This implies that energy can be stored in certain spatial
domains. Clearly, the constant phase $\theta$ indicates the
appearance of a long range ordering in the system. We note that in
the superconductivity case, the uniform phase ($\nabla\theta=0$)
and a finite super-current lead to an infinite density of
carriers, indicating the condensation of electron pairs in the
momentum space and resulting in the infinite conductivity.

The above simple argument naturally leads to the question of
whether there are systems which can realize the above features. We
found that these features can appear in a number of different
systems. The previously considered acoustic propagation in bubbly
water\cite{cjp} is one of such systems, as well as acoustic
propagation in water with air-filled cylinders\cite{Ye}, and the
electric diople system.

For all the systems mentioned, in the linear approximation the
governing equation can be reduced to\cite{Ye,dipole} \BA
\frac{d^2}{dt^2}\psi_i &+& \omega_0^2\psi_i + \gamma\frac{d}{d
t}\psi_i = F_i\nonumber\\ &+& \sum_{j=1,j\neq i}^N \left.C
\frac{\ddot{\psi_j}}{|\vec{r}_i-\vec{r}_j|^{(d-1)/2}}\right|_{t-|\vec{r}_i-\vec{r}_j|/c},
\label{eq:4} \EA where $F_i$ represents the external stimulation,
$C$ is the coupling constant, $d$ denotes the dimension, and $c$
is the wave speed. Strictly, for the 2D ($d=2$) acoustic case, the
propagator on the RHS would often be in a form of the zero-th
order Hankel function of the first kind\cite{Ye}. The quantity
$\psi$ refers to either the oscillation of the dipoles in electric
systems and or to the vibration of the air-filled bodies in the
acoustic situation, and $\omega_0$ refers to the corresponding
natural frequency, and $\gamma$ the possible damping effects due
to such as radiation. The parameters will depend on specific
models considered. For acoustic scattering in bubbly water, for
example, $C\sim a$, $\gamma \sim \omega_0$, and $\omega \sim
\frac{1}{a}\sqrt{p_0}$. Here $a$ is the radius of air-bubbles and
$p_0$ is the ambient pressure. In this case, $\psi$ refers to the
radial pulsating of the bubbles. The parameters can be found in
\cite{cjp} and references therein. For the electric dipole
systems, $C=-\frac{q^2\mu_0}{2\pi m}$ with $q$ being the charge
and $m$ the mass of the charge, and $\psi$ is the electric dipole.

Eq.~(\ref{eq:4}) reveals that when it is excited by the incoming
wave, either of electronmagnetic or of acoustic nature, the
constituent will radiate or scatter waves. The radiated or
scattered waves will be again re-radiated or scattered by other
constituents, a process repeated to establish an infinite
recursive pattern of multiple interactions. We note here that
Eq.~(\ref{eq:4}) can in fact also describe many other systems not
being considered here.

To explore the general properties of the model in
Eq.~(\ref{eq:4}). We solve it numerically in the frequency domain.
The generic set up is as follows. There are $N$ bodies randomly
distributed in a system, and they are excited by a point source
located roughly in the middle of these bodies. We further assume
that all elements in the system are identical. We write
$\psi_i=A_ie^{i\theta_i}$, where we dropped the time factor
$e^{-i\omega t}$.

For each phase $\theta_i$, by analogy with \cite{cjp} we define a
phase vector such that $\vec{v}_i=\cos\theta_i\vec{e}_x +
\sin\theta_i\vec{e}_y.$ The phase vectors would indicate the
degree the coherence of the vibration behavior. We have done a few
general numerical calculations for Eq.~(\ref{eq:4}). The
significant discoveries we found can be summarized as follows. For
two or three dimensions, the energy is localized near the source
within a range of frequencies slightly above the natural frequency
for sufficiently large coupling constants and number densities of
the constituents. When the energy is localized in the system, all
the constituents oscillates in phase, i.~e. the phase $\theta$
becomes constant, revealing a long range ordering in the system.

Here we would like to show one example. Randomly put $N$ identical
electric dipoles on the $x-y$ plane, inside a box of side length
$L$. All the dipoles points to the positive $z$-axis. Assume that
the averaged distance between dipoles is $d$; thus the number
density of the dipoles is $n \sim 1/d^2$. In the computation, the
damping factor $\gamma$ can be adjusted to reflect various
situations. A transmitting source is put at the center of the
dipole cloud. As this is the first step in our research, we
simplify further that all the dipoles can only oscillate along
their axes. The dipoles will oscillate vertically in response to
the source and as well as the radiated waves from other dipoles,
i.~e. all the dipoles are coupled via the radiated waves. In the
computation, all frequencies are scaled by the natural frequency
of the dipole. The coupling constant $C$ and damping $\gamma$ can
be varied. We study the behavior of the aforementioned phase
vectors and the energy distribution in the system; it is easy to
see that the energy can be represented by the oscillation
amplitude of the dipoles.

\begin{figure}
\begin{center}
\epsfxsize=5in\epsffile{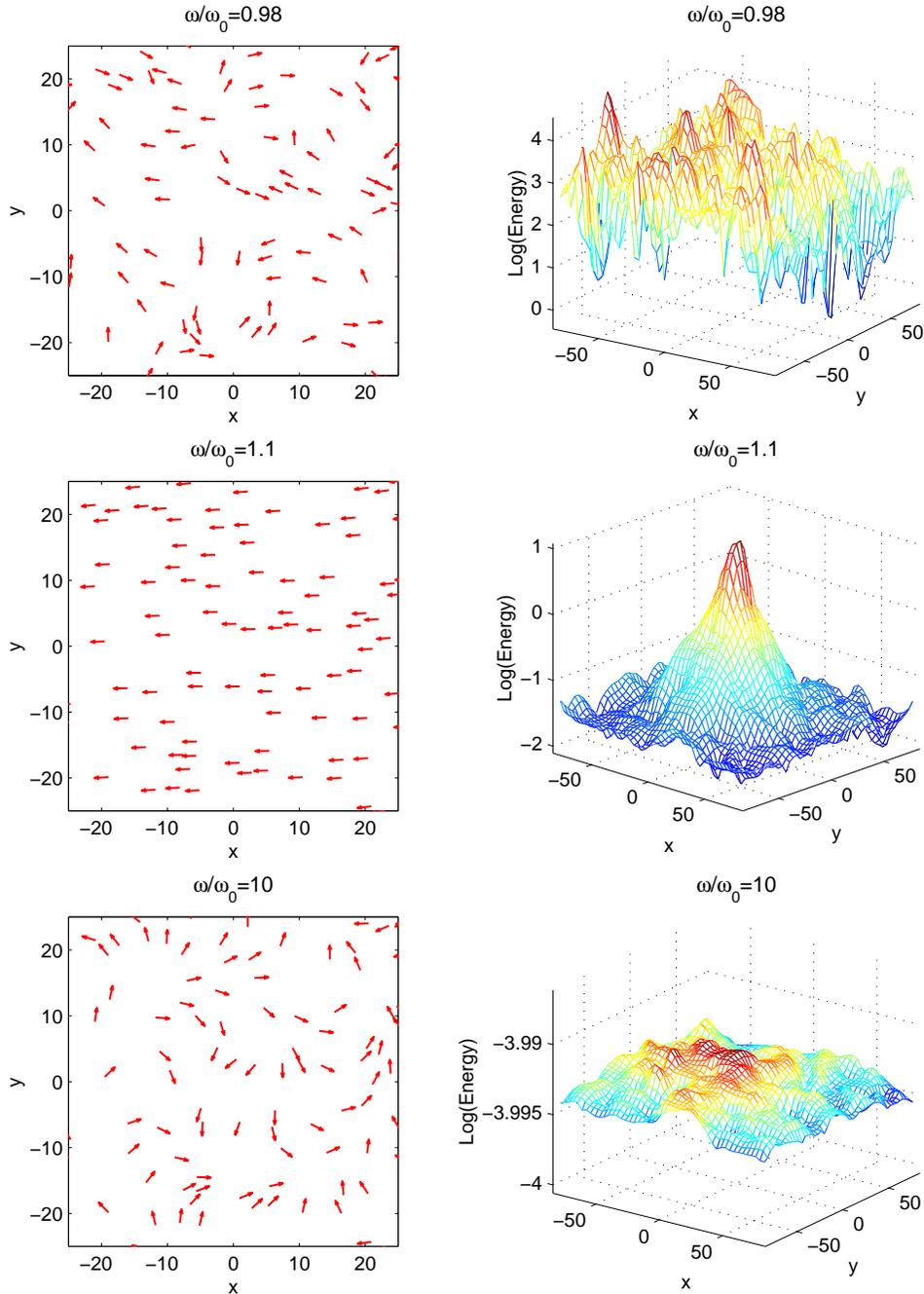} \caption{Energy distribution and
diagrams for the phase vectors in a 2D random configuration of
electric dipoles. Right: Energy distribution; the geometrical
factor ($1/r$) has been dropped out. Left: the phase diagrams for
the phase vectors defined in the text.} \label{fig1}
\end{center}
\end{figure}

Fig.~\ref{fig1} illustrates our findings for one arbitrary random
configuration. For low frequencies, e.~g. at $\omega/\omega_0 =
0.98$, the phase vectors points randomly. The distribution of the
energy spreads. When increasing the frequency to certain values
slightly above the natural frequency, all phase vectors tend to
point to a uniform direction, indicating a new phase state.
Meanwhile the energy is centered at the site of the transmitting
source. This is shown by the case at $\omega/\omega_0 = 1.1$. When
the frequency is increased further, the phase ordering disappears,
and the energy distribution becomes extended again, as shown by
the case with $\omega/\omega_0 =10$. These features are valid for
a range of the coupling constant $C$ and the damping rate. The
parameters used in the present specific example are: $C=0.1$,
$n=0.032$, and $\gamma = 0.0001$. We take $N$ as 900; in Fig.~1,
however, only a fraction of the system is shown in order to expose
the features in the most explicit way. The features tend to
diminish as either the couple constant or the number density is
reduced to a certain value. When the damping rate $\gamma$ is
increased to a large value, the phase ordering will then be
degraded. The phase ordering in the present system has a clear
physical meaning. That is, it reflects that all elements in the
many body system oscillate or vibrate coherently. We stress that
for simplicity we have choosen the 2D arrangement of the dipoles
in a 3D system. Strictly speaking, due to radiation into the third
dimension, waves cannot be stored permanently in the medium. The
energy will gradually escape. We have also performed calculation
for a 3D random configuration of the dipoles. The similar results
hold.

In summary, we have exhibited a new phase state possibly
observable in a class of many body systems. It is shown that
energy can be stored coherently in random media. The discovery
reported here may provide insight to the long standing problem of
the Anderson localization of waves in disordered media.

\noindent The work received support from National Science Council
through the grants NSC-89-2112-M008-008 and NSC-89-2611-M008-002.

\newpage

\twocolumn[\hsize\textwidth\columnwidth\hsize\csname@twocolumnfalse\endcsname

\input epsf.tex

]

\end{document}